\documentclass[epj,referee]{svjour}
\usepackage[T1]{fontenc}
\usepackage{graphicx}
\usepackage{amsfonts}
\usepackage{amssymb}
\usepackage{subfigure}
\usepackage{setspace}
\usepackage{mathtools}
\usepackage{latexsym}
\usepackage{graphics}

\begin{document}
\title{Oscillation patterns in active emulsion networks}
\author{
Shashi Thutupalli
\thanks{\emph{Present address:} Princeton University, Princeton, NJ, USA}
\and Stephan Herminghaus
}
\mail{shashi@princeton.edu}
\institute{Max Planck Institute for Dynamics and Self-organization, G\"ottingen, Germany}

\date{Received: date / Revised version: date}

\abstract{
We study water-in-oil emulsion droplets, running the Belousov-Zhabotinsky reaction, that form a new type of active matter unit. These droplets, stabilised by surfactants dispersed in the oil medium, are capable of internal chemical oscillations and also self-propulsion due to dynamic interfacial instabilities that result from the chemical reactions. The chemical oscillations can couple via the exchange of activator and inhibitor type of reaction intermediates across the droplets under precise conditions of surfactant bilayer formation between the droplets. Here we present the synchronization behaviour of networks of such chemical oscillators and show that the resulting dynamics depend on the network topology. Further, we demonstrate that the motion of droplets can be synchronized with the chemical oscillations inside the droplets, leading to exciting possibilities in future studies of active matter.
} 

\maketitle
\section{Introduction}

There has been rapidly growing interest recently in so-called active matter, which refers to open (soft) matter systems exhibiting complex dynamics and collective behavior reminiscent of living organisms. Spontaneous oscillations and self-sustained motion \cite{ramaswamy_mechanics_2010,kruse_oscillations_2005} represent the simplest examples of such complex dynamics and hence are ideally suited for a theoretical analysis of the interactions and dynamic properties of a complex active system. Systems such as those undergoing catalytic reactions at interfaces, the autocatalytic Belousov-Zhabotinsky (BZ) reaction, social amoeba under stress, populations of fireflies or the cells of heart muscle tissue can all display spatio-temporal oscillatory behaviors following similar patterns, which may be modelled by means of reaction-diffusion dynamics or suitable mean-field approaches \cite{kuramoto,Strogatz2000}. Self-propelling entities such as motile bacteria, sperm, birds, and fish, or analogous physical systems such as active emulsions have been found to exhibit remarkable similarities in their collective behavior as well \cite{dos_santos_free-running_1995,sumino_self-running_2005,linke_self-propelled_2006,Howse2007,thutupalli_njp_2011}. There has been tremendous theoretical and experimental progress towards the understanding of such systems in terms of the dynamics of oscillators and motile particles. However, most of the treatment of these phenomena has been in isolation \cite{ramaswamy_mechanics_2010,kuramoto} and in most of the studies of the collective behavior of active matter, the individual unit has typically been abstracted to be either a point like self propelled particle or a simple phase oscillator \cite{kuramoto,Strogatz2000,thutupalli_njp_2011,toner_hydrodynamics_2005,bhattacharya_collective_2010,Schaller2010,guttal_social_2010,romanczuk_collective_2009}. 

While these studies have been very successful in understanding a range of collective phenomena and synchronization, this is clearly not the complete picture, as is obvious in many biological processes. For instance, during the chemotactic self-organization of amoeba and in cellular organization during embryogenesis, the internal dynamics are entwined with the macroscopic order that emerges due to cell motility. These internal dynamics of an individual unit, most often, manifest as biochemical oscillations, linked together with the motility of the unit. Even in simple physical systems, it has been observed that there is spontaneous symmetry breaking and emergence of unexpected complex behavior when the internal degrees of freedom of coupled non-equilibrium entities are taken into account \cite{danTanakaPRL,abramsChimeraPRL,chimeraMetronomes,erikChimerasChaos}. Therefore, it may be expected that a range of rich collective phenomena in active matter might open up when the internal dynamics such as oscillations are studied together with the resultant (or already existing) dynamics such as motility.

Here, we introduce a system that can potentially be a simple table-top experiment to study the interplay between the internal dynamics of an individual unit and its motion, and hence the collective behavior of the whole system. Specifically, we enclose the Belousov-Zhabotinsky reaction in emulsion droplets of a few tens to hundreds of microns in diameter to form chemical oscillators. These chemical oscillators can also sustain self-propelled motion with respect to the surrounding medium due to interfacial instabilities and Marangoni stresses.

\section{Experimental Techniques}

\begin{figure*}[ht]
\includegraphics[width=\textwidth]{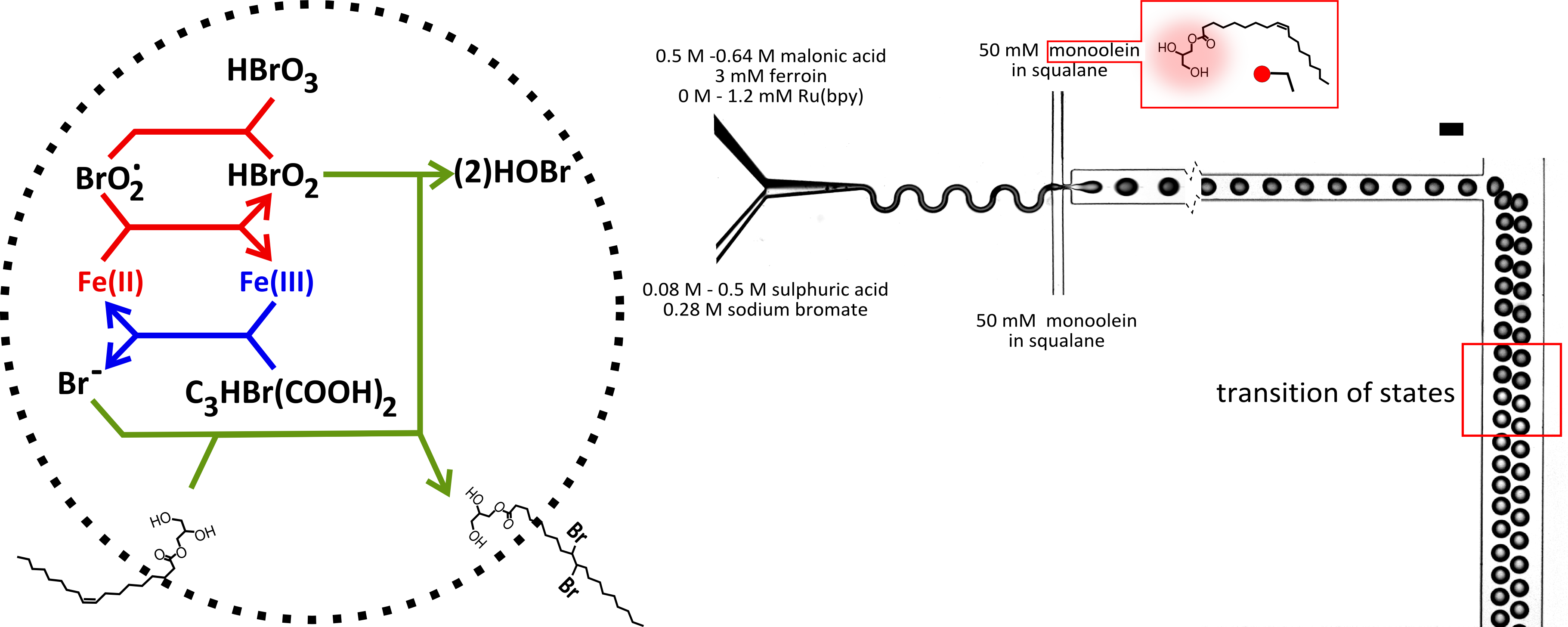}
\caption{Production of monodisperse oscillator droplets. Left: The BZ reaction consists of two loops (i) an autocatalytic and (ii) an inhibitory cycle. The reaction state can be visualised by the colour of the ferroin catalyst. In our setting, an additional reaction with the unsaturated mono-olein surfactant occurs as shown. Right: The contents of the BZ reaction are mixed on the microfluidic chip to prevent any pre-reaction. Seen through an optical 480/20 nm notch filter, the transition from red colour to the blue colour of the BZ reaction is seen by the change in the brightness of the droplet.}
\label{fig:oscProduction}
\end{figure*}

Our chemical oscillators are made from incorporating the BZ reaction mixture in aqueous droplets in an external oil phase of squalane contaning mono-olein as surfactant.  The droplets for these experiments were generated using a flow focussing channel geometry in a PDMS microfluidic chip as shown in Fig. \ref{fig:oscProduction} \cite{RepProgPhys2012}. In order to prevent any pre-reaction and the formation of unwanted gaseous bubbles of carbon-di-oxide, the BZ reaction mixture is separated into two parts and they are combined on chip. The two parts are created in stock with concentrations as follows: (i) 500 mM sulphuric acid ($\rm H_2SO_4$) and 280 mM sodium bromate ($\rm NaBrO_3$) (ii) 300 - 800 mM malonic acid ($\rm C_3H_4O_4$) and 3 mM ferroin ($C_{36}H_{24}FeN_6O_4S$). The concentration of the mono-olein in the squalane ranges between 25 - 100 mM. 

As can be seen from the left panel of Fig. \ref{fig:oscProduction}, the BZ reaction consists of an autocatalytic cycle in which $\rm HBrO_2$ catalyses its own production via the reduction of the ferroin catalyst, which changes its colour rapidly from red to blue in response. This is when the inhibitory cycle proceeds, leading to a slow production of bromine which quenches the autocatalysis. The effect of this is a gradual change of the catalyst back from the blue colour to red. In our case, an additional side-reaction occurs due to the addition of mono-olein as a surfactant, which is used to stabilize the droplets against coalescence. Since the surfactant has an unsaturated hydrocarbon chain as shown, some of the bromine that is produced in the inhibitory cycle rapidly reacts with the unsaturated bond. As we will discuss in more detail below, this 'trapping' of the bromine by the surfactant significantly affects the coupling between droplet oscillators in our setup.

\begin{figure}
\centering
\includegraphics[width=\columnwidth]{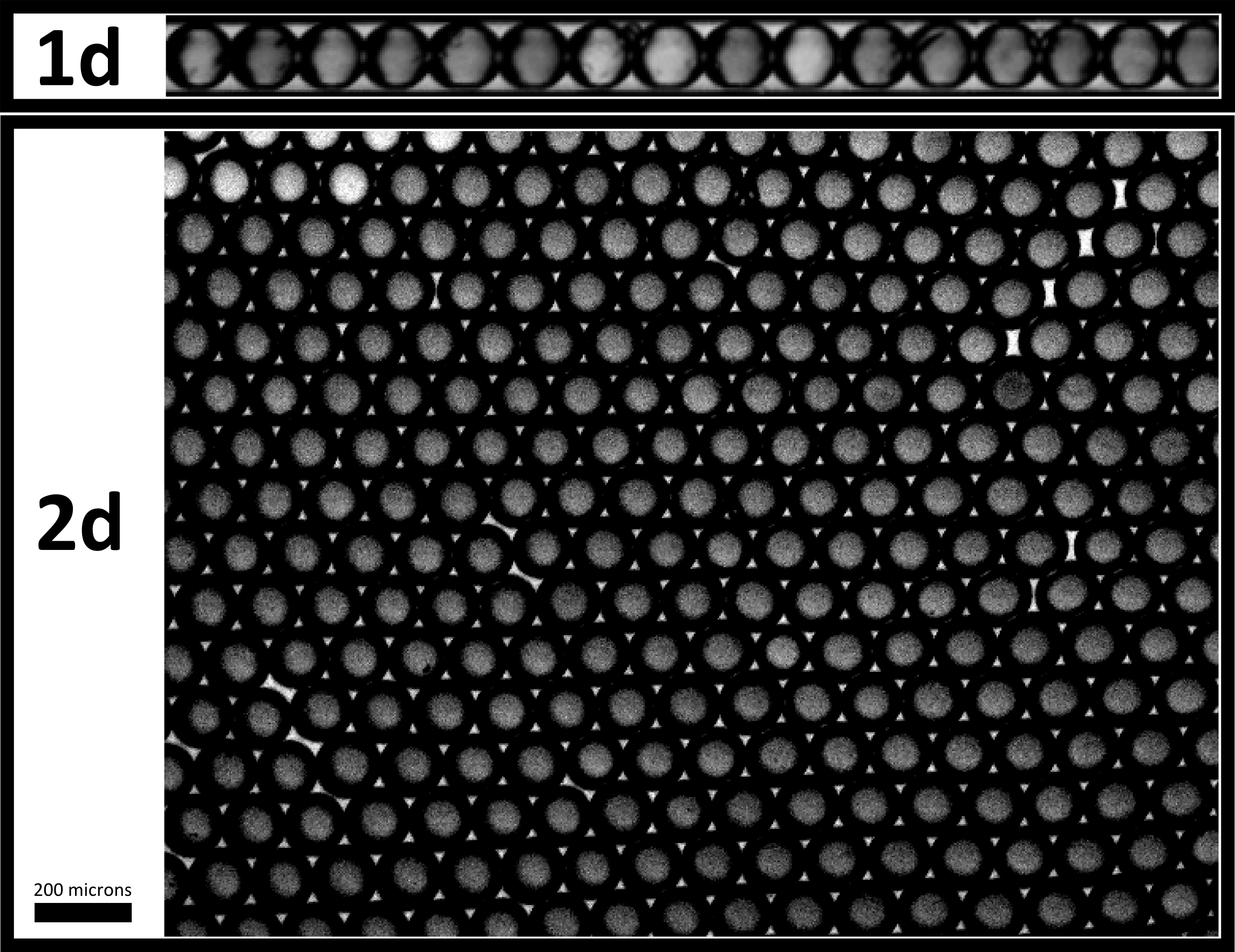}
\caption{Storage of droplet oscillations in one and two dimensional confinement for observation of their dynamics.}
\label{fig:oscStorage}
\end{figure}

\begin{figure}
\includegraphics[width=\columnwidth]{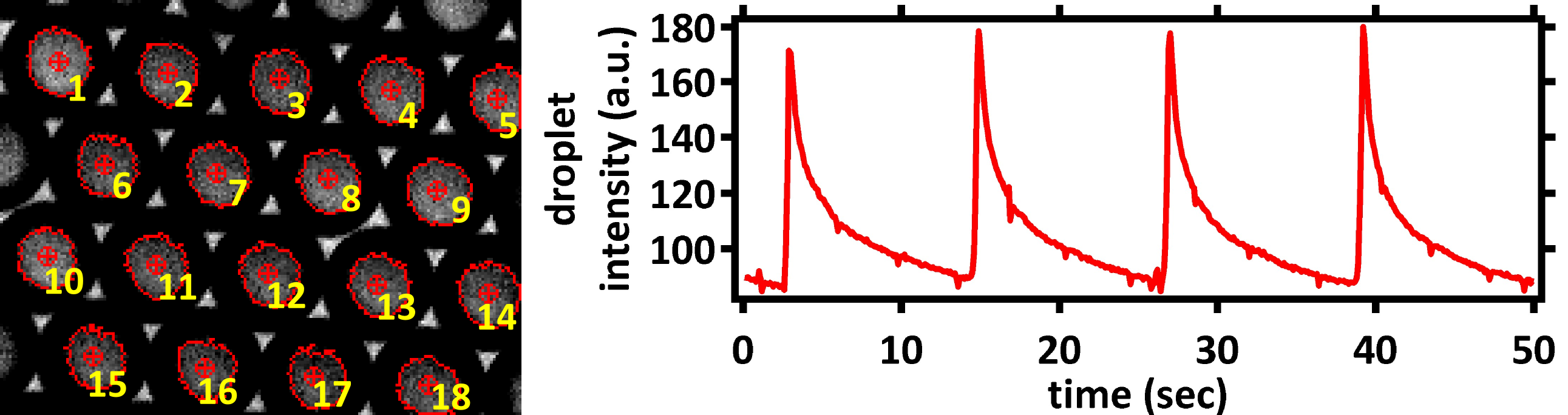}
\caption{Image processing to identify the droplet oscillators and record their dynamics.}
\label{fig:oscProcessing}
\end{figure}

The droplet oscillators are stored as a monolayer in either a one-dimensional (1d) or two-dimensional (2d) arrangement, as shown in Fig. \ref{fig:oscStorage}. The 1d array is created within a glass capillary with a square cross-section of inner width 100 $\rm \mu$m and outer width 135 $\rm \mu$m (Hilgenberg GmbH, Germany). The inner walls of the capillary are hydrophobised using a coat of commerically available hydrophobising agent 'Nano-protect' (W5 Carcare). The 2d array is created between two similarly hydrophobised glass slides with a PDMS spacer. The reaction dynamics are recorded by video microscopy on an inverted microscope (Olympus IX 81) through an appropriate optical filter. As described earlier, the BZ reaction dynamics can be followed by the colour of the catalyst as it changes from red to blue and vice versa. The optical filter we chose is therefore a notch filter of 480/20 nm wavelength such that the red colour of the catalyst has less transmittance through the filter. Droplets are identified from the recorded images using Image-Pro Plus (Media Cybernetics) as shown in Fig. \ref{fig:oscProcessing} (left panel). The BZ oscillations within the droplet are then identified by measuring the mean intensity value of the droplet. They are recorded as  traces similar to those of a relaxation oscillator as seen in the right panel of Fig. \ref{fig:oscProcessing}. The sudden rise of intensity corresponds to the autocatalytic cycle of the BZ reaction with the catalyst changing from red to blue colour, and the gradual fall in intensity corresponds to the release of bromine in the reaction, thus changing the catalyst colour back to red. Further analysis on the obtained data as described in the results section is done using MATLAB.

\section{Results and Discussion}

\subsection{Isolated BZ oscillators}

\begin{figure}[ht]
\centering
\includegraphics[width = \columnwidth]{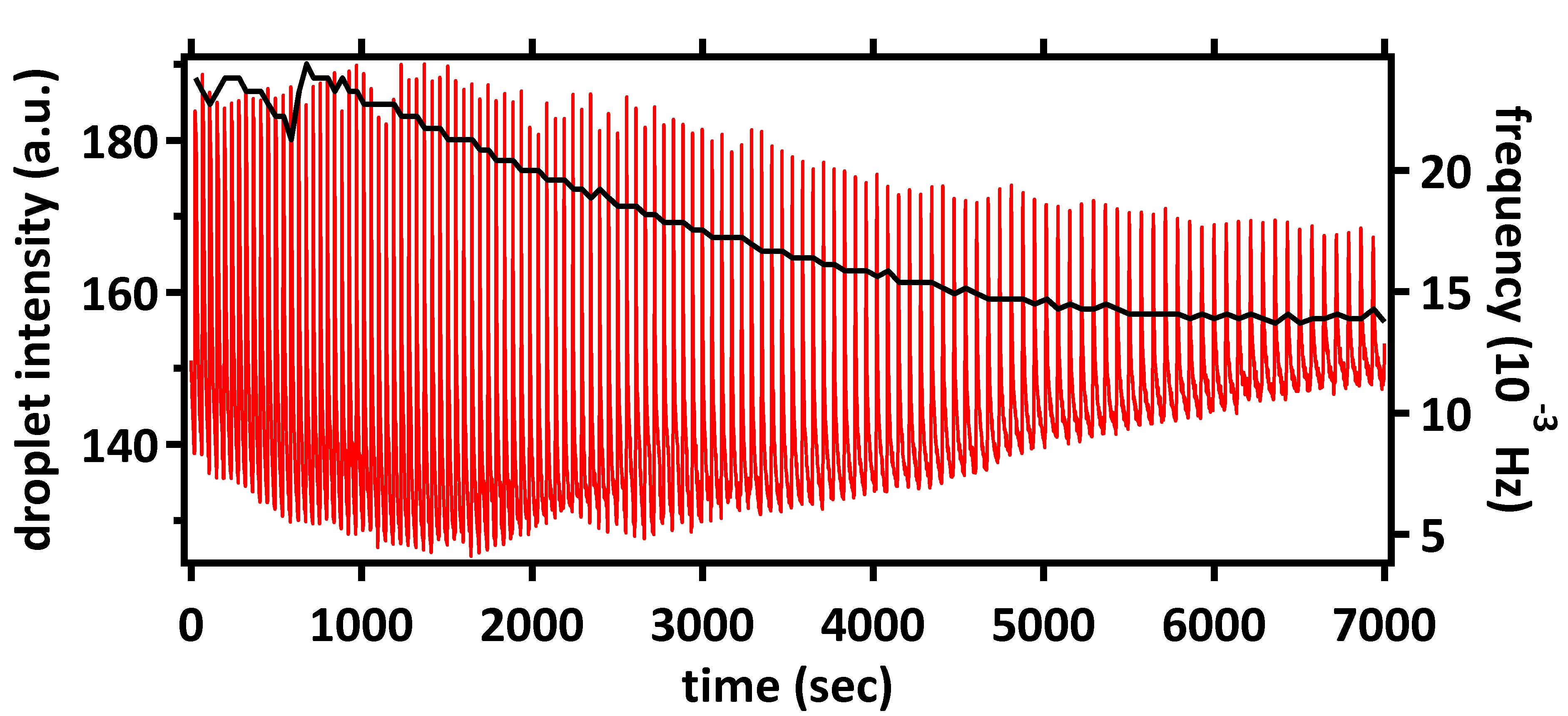}
\caption{BZ oscillations in a closed reactor. The oscillation trace and its corresponding frequency (black) are plotted as a function of time. 
\label{fig:oscIsolated}}
\end{figure}

The BZ oscillators described here are a closed reactor system i.e. there is no cycling of reactants and final products during the course of the experiment. Therefore all the reactants and the resultant products remain within the droplet, except some outflux of promoter and inhibitor into the oil phase. As a result, the nature of the oscillations gradually changes with time. As soon as the experiment is started, the oscillation is set by the initial reaction conditions and the amplitude of the oscillations and the frequency change as time proceeds. This can be seen in Fig. \ref{fig:oscIsolated}. As time proceeds, the amplitude of the oscillations reduces significantly, until they die out completely. The frequency of the oscillation, shown in black, is gradually reduced as well. In a sense, however, the BZ oscillators provide their own clock, and the oscillators suggest themselves as a time normal of the experiment if applicable. In any case, we did not observe any qualitative change of the behaviour of our system as time proceeded and droplet oscillations slowed down.

\begin{figure}[ht]
\centering
\includegraphics[width = \columnwidth]{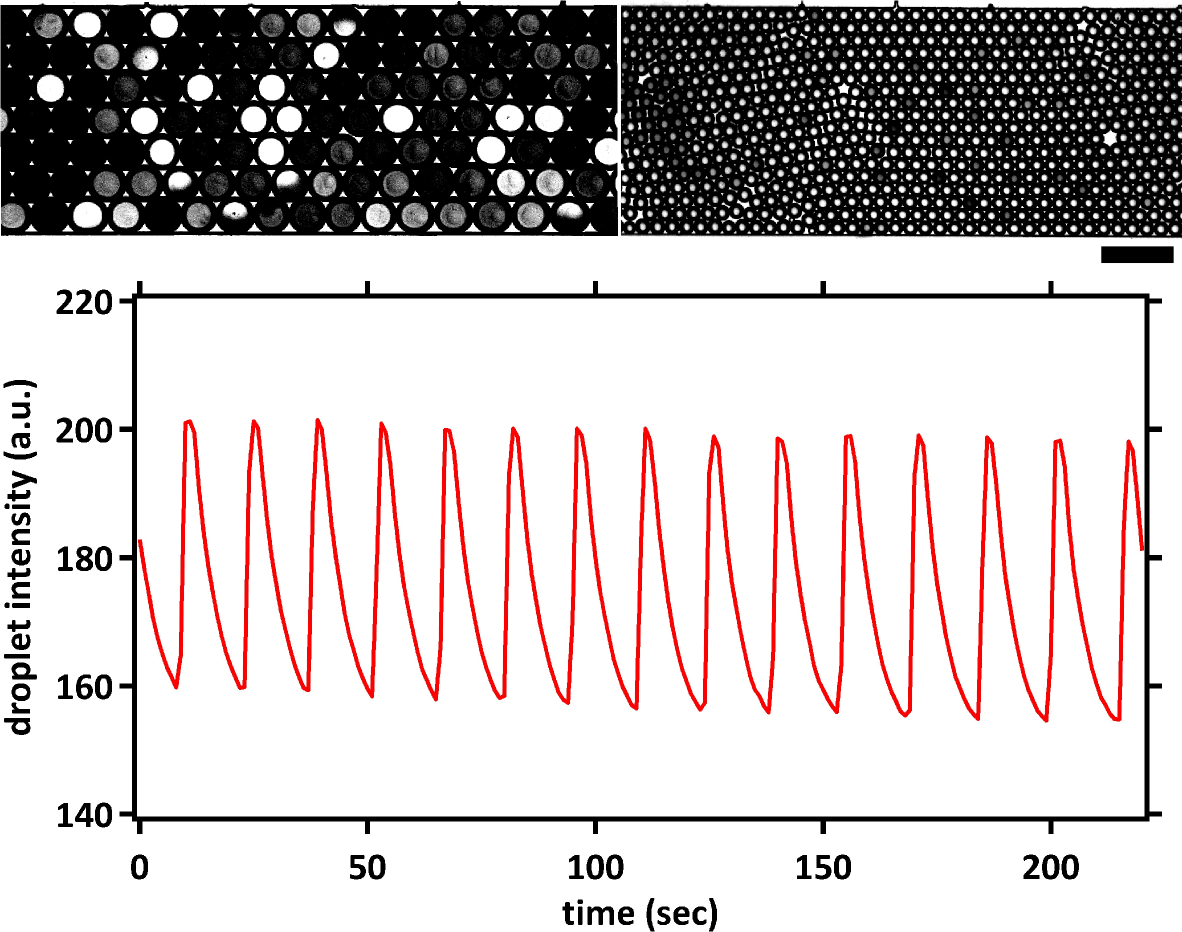}
\caption{Droplet oscillators in a hexagonal packing geometry within a PDMS microchannel. Top: The different intensities of the droplets show the different BZ reaction states within each droplet for two different droplet sizes. Each droplet acts like an isolated individual oscillator without any coupling with its neighbours. Image contrast is enhanced for better visualization. Scale bar is 150 microns. Bottom: The intensity trace for a single droplet is shown as a function of time. The constancy of the frequency and amplitude of the oscillations can be clearly seen.
\label{fig:oscNoMembranes} }
\end{figure}

The BZ oscillators are suspended in an oil phase consisting of squalane, with mono-olein at concentrations well above the critical micelle concentration (CMC). The mono-olein serves two purposes. First, it forms dense surfactant layers at the oil/water interface, which  readily form bilayer membranes if brought into close proximity.  Second, the C=C double bond in the mono-olein molecule acts as an efficient scavenger for bromine, since the latter rapidly reacts with this site. The oil phase is thus expected to efficiently suppress coupling between neighboring droplets, which is mediated by the exitatory and the inhibitory species. That this is indeed the case can be seen in Fig. \ref{fig:oscNoMembranes} which shows a two dimensional hexagonal packing of BZ oscillators. The spherical shape of the droplets in the packing clearly shows that there is oil between the droplets and that bilayer membranes have not yet formed  \cite{thutupalli_SM_2011}. The oscillation trace of a single oscillator is shown in the lower panel of Fig. \ref{fig:oscNoMembranes}. We note that in our experiments we do not see a systematic dependance of the oscillation frequency on the droplet size. When such isolated droplets are close to each other, in spite of the fact that the diffusion of the excitatory and inhibitory species can indeed cause coupling between droplet, they are observed to be uncoupled. This is due to the fact that, as we discussed before, they might be trapped via reaction with the surfactant molecules.

\begin{figure}[ht]
\centering
\includegraphics[width = \columnwidth]{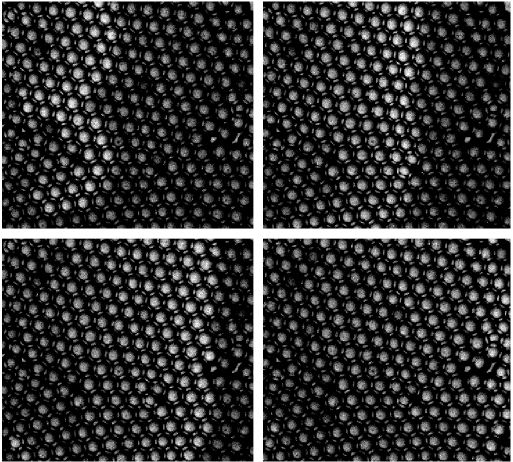}
\caption{The formation of travelling waves when bilayer membranes are formed between oscillator droplets. Each image is 5 seconds apart. The droplet diameter is 30 microns.}
\label{fig:oscMembraneSpiral}
\end{figure}

However, bilayer membranes form spontaneously between the droplets \cite{thutupalli_SM_2011}, and this happens in the case of the droplet oscillators too. As soon as bilayers form between droplets, their interfaces touch each other very closely such that the droplets are not perfectly spherical anymore. Consequently, the packing fraction increases as seen in Fig. \ref{fig:oscMembraneSpiral}, where the gaps between the droplets due to oil that existed before bilayer formation are now reduced significantly. Once the bilayers are formed, completely different oscillatory dynamics are seen. Previously, we demonstrated that oscillator coupling can be initiated by the formation of a bilayer membrane between the oscillator droplets \cite{thutupalli_SM_2011}. Often, waves of synchronised activity such as travelling waves are seen as in Fig. \ref{fig:oscMembraneSpiral}. Therefore we see a switch from the individual to collective dynamics of the oscillators when bilayer networks are formed. We discuss this aspect in greater detail in the next section.

\subsection{Synchronization patterns}

Patterns such as pacemaker driven target waves, travelling waves and spirals are most commonly seen in large assemblies of coupled oscillators. The BZ droplet oscillators, connected by bilayer membranes, also give rise to  a rich variety of collective dynamics. In the present section, we discuss the various patterns that emerge in connected networks of oscillators and the dependance on the network topology of the type of behaviour that emerges. The discreteness of the droplet oscillators allows us to clearly identify trigger locations within the networks. All the following experiments are done with a BZ reaction mixture as described in the experimental section, with a malonic acid concentration of 500 mM. 

\begin{figure}[ht]
\includegraphics[width = \columnwidth]{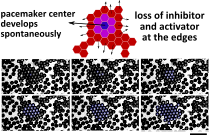}
\caption{Formation of a target pattern in a 'island' or 'peninsula' type of droplet network. Top: A schematic of a hexagonal arrangement of droplets which form an 'island'. At the outer edges of the structure, the excitatory and the inhibitory components are lost to the oil phase (shown by the arrows), while in the center, they are concentrated leading to the formation of pacemaker center. Bottom: A target pattern develops within a 'peninsula' of droplet oscillators. The excitation and wave pattern are shown in blue for easy visualization. The scale bar is 500 microns.
\label{fig:oscPatternPF} }
\end{figure}

First, we discuss the formation of target waves. These are characterised by a pacemaker core which periodically tiggers excitatory waves that spread from the core center outward. In our system, we observe that pacemakers spontaneously emerge in the center of connected droplet 'islands' or 'peninsulas'. An 'island' is comprised of connected droplets as shown in the left panel of Fig. \ref{fig:oscPatternPF} with the outer edges of the 'island' open to the mono-olein filled oil phase. A 'peninsula' is a similar structure and is connected by a narrow bridge of one or two droplets to a neighbouring 'island'. As we discussed before, the inhibitory (bromine) and the excitatory ($\rm BrO^{\cdot}_2$) components of the BZ reaction readily diffuse into the external oil phase, where they are trapped by the surfactant. Therefore, at the edges of the island, the oscillatory droplets lose their inhibitory and excitatory components to the external oil phase. However, at the center of the island, the concentration of the BZ coupling species increases since they come in from all sides. Depending on the relative concentrations of the inbitory and excitatory components, the center droplet can therefore either 'turn off' (i.e. oscillations are inhibited) or trigger an oscillation. For a malonic acid concentration of 500 mM together with the other concentrations as described in the experimental section, we find that an oscillation is triggered in the central droplet as can be seen in the right panel of Fig. \ref{fig:oscPatternPF}. This trigger from the central droplet then propagates outward as a target wave throughout the 'island'. If it is a 'peninsula' the connecting bridge can couple the wave to the neighbouring 'island' as well. This pattern then repeats periodically. We find that droplets which are not connected via a bilayer to the synchronously oscillating cluster are quite likely to not oscillate at all. This represents an instance of quorum sensing, as reported before in a similar system \cite{Showalter_QS_PRL}.                

\begin{figure}[ht]
\includegraphics[width = \columnwidth]{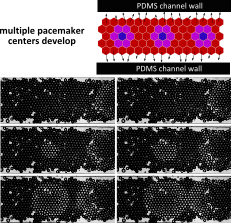}
\caption{Formation of target waves with multiple pacemaker centers in a oscillator network confined in a PDMS microchannel. Top: Schematic of the oscillators in a PDMS channel. PDMS, in addition to the oil phase, absorbs the excitatory and inhibitory components of the BZ at the edges (shown by the arrows). Bottom: Two target wave patterns seen in the channel. One target wave is forming at a pacemaker center clearly visible. The center of target pattern coming in from the right is not visible. \label{fig:oscPatternPC} }
\end{figure}

Next, we sought if we could induce the pacemaker patterns by confining the oscillator droplets within a channel made of PDMS. In such a scenario as shown in Fig. \ref{fig:oscPatternPC}, the PDMS walls of the channel, in addition to the oil phase, act as sinks for the BZ reaction species. Therefore, we expect that along the length of the channel, multiple pacemaker centers form, each triggering target waves. That this is indeed the case, can be seen in the right panel of Fig. \ref{fig:oscPatternPC}. The triggering of a wave from a core can be seen in the image sequence shown. In addition, a wave can be seen coming in from the right of the images, clearly triggered by a pacemaker upstream in the channel. Indeed there were also waves coming in the from the left side, but are not shown here. 

\begin{figure}[ht]
\centering
\includegraphics[width = \columnwidth]{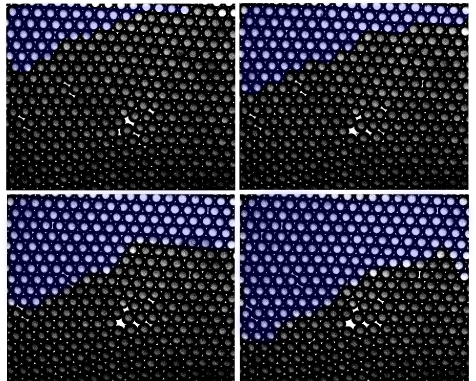}
\caption{Travelling waves are formed in large densely connected oscillator networks. Each image is spaced 5 seconds apart. The excitation is coloured blue for easy visualization. \label{fig:oscPatternTW} }
\end{figure}

Next, for the same concentrations, we looked at a very large network of hexagonally packed droplet oscillators, such that the edges are too far away from the cores to have a significant impact. This is shown in Fig. \ref{fig:oscPatternTW}. In such a case, we see the spontaneous emergence of travelling waves across the network. Indeed, it may be expected that since the BZ concentrations are rather uniform over the network, a random trigger in one of the oscillators can set off a cascading wave of activity, which repeats periodically. However, we were not able in this scenario to find the precise conditions and locations at which the waves were triggered. 

\begin{figure}[ht]
\centering
\includegraphics[width = \columnwidth]{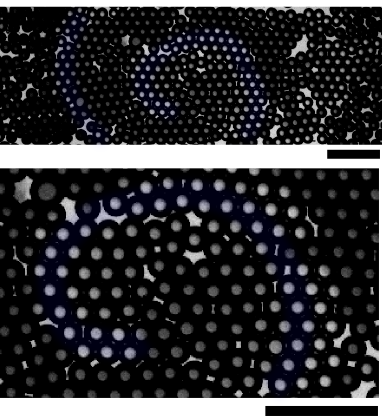}
\caption{Formation of spirals in oscillator networks. Top: A spiral can be seen clearly in the oscillator population. Bottom: Image of the network at higher magnification reavealing the lack of perfect local connectivity for each droplet. Scale bar is 150 microns.
\label{fig:oscPatternSpiral} }
\end{figure}

Spiral waves formed in our experiments, when the hexagonal packing was not perfect such that not every oscillator is coupled to 6 nearest neighbours. Yet, the networks were not so sparse as to form 'islands' or peninsulas', where target waves were predominant. An instance of a spiral is seen in Fig. \ref{fig:oscPatternSpiral}. A spiral wave can be clearly seen among the oscillator population. A closer look into the network reveals that the local network of each oscillator is not complete according to the 2-dimensional hexagonal packing. The lack of local connections creates a refractory effect on the excitatory wave due to the different speeds it travels at, in the different directions. This causes the wave to turn and eventually forms a spiral or other rotary patterns depending on the exact topology of the network. Such defects are well known to provide cores for spiral waves also in other systems \cite{DictySpiral1974,LutherHeartSpiral2011}

\begin{figure}[ht]
\centering
\includegraphics[width = \columnwidth]{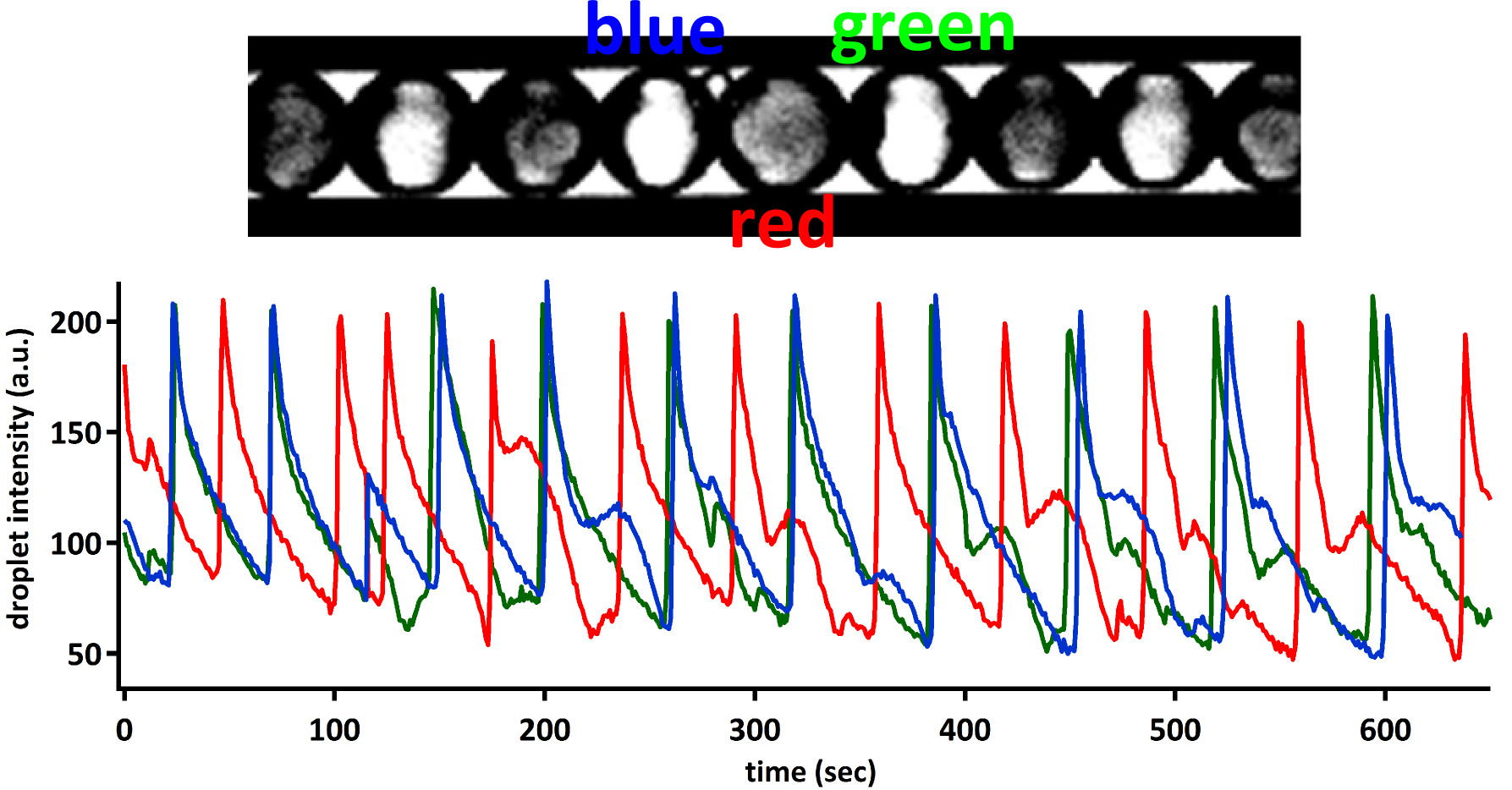}
\caption{Antiphase pattern in a 1 dimensional oscillator network. Top: Droplet oscillators in a glass microcapillary. Each droplet pair is connected by bilayer membranes. The droplet diameter is 100 microns Bottom: Time trace of the droplet oscillations shown for the three droplets in the center of the top image. The red, blue and green traces correspond to the droplets marked as shown. 
\label{fig:oscPatternAP} }
\end{figure}

Finally, we note that the effect of the bilayer membrane is not just to easily pass the various species of the BZ reaction from one oscillator to another, such that a discrimination of the 'individual oscillator' is simply lost. In such a case, the behaviour of the oscillator network can be considered to be the same as that of the BZ reaction in the bulk, in a 2d homogeneous planar system. However, more complex relationships between neighbouring oscillators connected by a bilayer are seen. As we mentioned before, both the excitatory and inhibitory components of the BZ reaction can traverse the bilayer. The formation of the target patterns as described before indicated that the BZ mixture used in our experiments is excitatory i.e. the excitatory coupling wins over the inhibitory coupling in determining the state of the coupled oscillators, leading to wave like patterns as we have seen so far. However, it has been reported in literature \cite{Toiya2008,Toiya2010} that due to inhibitory coupling between BZ oscillators, it is possible to generate patterns that strongly differ from wave-like patterns. We increased the concentration of malonic acid to 700 mM compared to the 500 mM used for the previous experiments. It is expected that increasing the concentration of the malonic acid results in a greater production of the inhibitor, bromine, as shown in the BZ reaction schematic in Fig. \ref{fig:oscProduction}, thus possibly leading to an inhibitory coupling effect. When the coupling is inhibitory i.e. non-excitatory, we expect that wave like patterns will not result. As anticipated, the increase in the concentration of Malonic acid resulted in a non-wave-like pattern as shown in Fig. \ref{fig:oscPatternAP} where every oscillator droplet is found to be in strict anti-phase with its neighbour. This can be understood to be a complicated interplay between the inhibitory and excitatory coupling. In fact it has been shown that the anti-phase state is an attractor for inhibitory coupling \cite{Toiya2008,Toiya2010}. However, in such models of inhibitory coupling, the interdroplet distance is quite large as compared to our experiments where the droplets are separated only by a nanometric membrane. This illustrates that the membrane between the oscillator droplets plays an important role in preserving the individual properties of each oscillator. These studies, though only demonstrative in the present work, must be performed in greater detail in order to quantify the various synchronization patterns and their relation to the network topology. In particular, a knowledge of the permeability of the membrane to the various coupling intermediates is crucial to have predictive control over the oscillator behaviour. 

\section{Summary and Outlook}

Active emulsions, consisting of chemical micro-oscillator droplets as presented here may provide a crucial first step towards the realization of active soft matter with complex dynamic functions. The Belousov-Zhabotinsky reaction used here has been studied as a paradigm system for dynamical and pattern forming systems for many years. In the present setting of using it within microfluidic emulsion droplets, qualitatively new phenomena emerge due to the interplay between the droplet network topologies and the type of coupling between the oscillators. As we have shown, the bilayer membranes, which form spontaneouly between adjacent droplets, play a crucial role in the coupling and synchronization dynamics. In combination with our previous results \cite{thutupalli_SM_2011}, this opens up the possibility to construct self-organizing dynamic soft matter systems. 

\begin{figure}[h]
\centering
\includegraphics[width = \columnwidth]{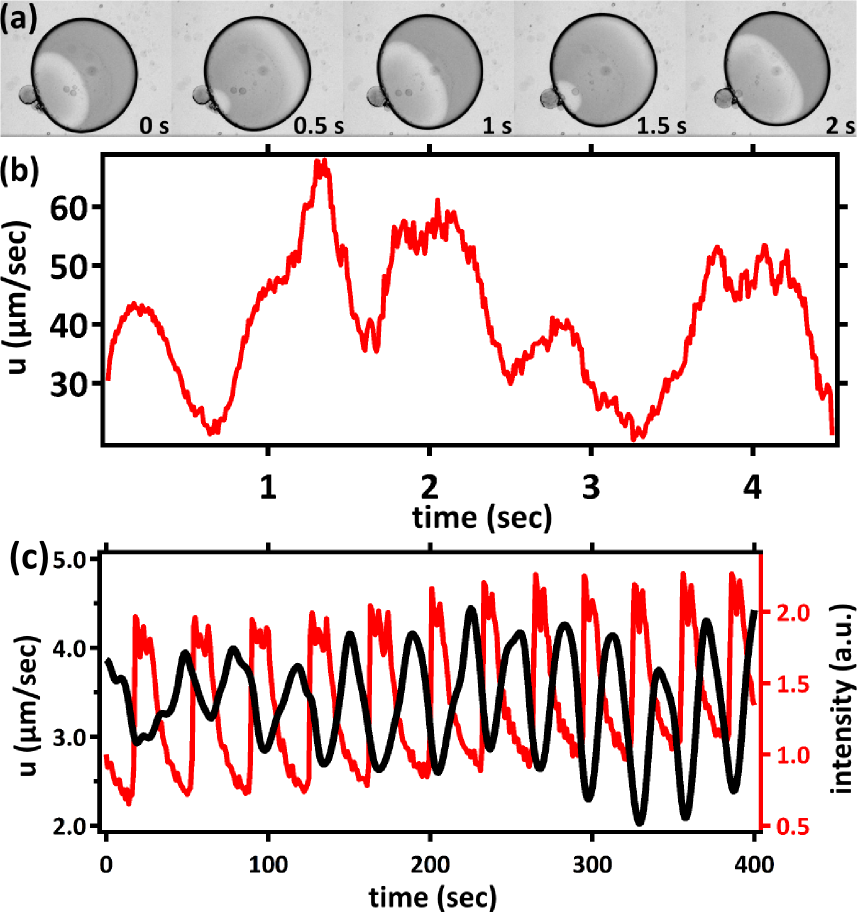}
\caption{A self propelled droplet with oscillating BZ chemical reaction taking place in the droplet. (a) Time snaps of a chemical wave within the droplet (diameter $\rm \sim600 microns$). The droplet moves in the direction of the wave propagation. (b) The speed of the above droplet shows roughly periodic oscillations corresponding to the BZ waves inside the droplet. (c) When the droplet size is reduced to $\rm \sim80 microns$ in diameter, the waves inside the droplets are supressed. In this case, the oscillations in the droplet speed (black trace) are perfectly synchronized with the optical transmission of the droplet, plotted in red. 
\label{fig:velocityOsc}}
\end{figure}

Also, we have shown previously \cite{thutupalli_njp_2011} that the BZ reaction intermediates react with the surfactant in the oil phase and also at the droplet interface creating artificial self propelled droplets. In Fig. \ref{fig:velocityOsc}, we demonstrate that the internal oscillations of the BZ reaction affect the speed of a self propelled droplet, similar to numerical predictions before \cite{Kitahata2011}.  Consequently, we can expect that the collective motion of droplets \cite{thutupalli_njp_2011} will be strongly  affected by their oscillatory state and mutual coupling. On the other hand, as we have seen, their local density affects the formation of bilayer membranes, and therefore acts back on the mutual coupling of individual droplet oscillators. This provides an exciting link to the rapidly evolving field of developmental evolution, which considers the possible back-action mechanisms of the emerging phenotype (i.e., the collective arrangement of cells) onto the genotype, i.e., the  microscopic (genetic) state of the cell \cite{evoDevo2002}. It is an interesting option to use emulsions containing chemical oscillator droplets such as BZ as a model for systems with such mutual interaction of different levels of integration. 

When mechanisms such as chemotaxis can be engineered into such systems, we envisage that many exciting possibilities might be opened in the field of active matter. For example, our results may provide a useful step to addressing some of the challenges in the design of artificial self organizing assemblies capable of achieving complex tasks \cite{Kolmakov2010}. Finally, we may develop well controlled experiments to understand how the internal degrees of freedom of collections of similar nonequilibrium units couple with their self-emergent mesoscopic order \cite{danTanakaPRL,thutupalliThesis}. 

\bibliographystyle{epj}
\bibliography{oscillatorsReferences}

\end{document}